\documentstyle[12pt]{article}
         \makeatletter
         \def\thefigure{\@arabic\c@figure}\def\fps@figure{tbp}
         \def\ftype@figure{1}\def\ext@figure{lof}
         \def\fnum@figure{\protect\footnotesize Fig.\ \thefigure}
         \def\thetable{\@arabic\c@table}
         \def\fps@table{tbp}\def\ftype@table{2}\def\ext@table{lot}
         \def\fnum@table{\protect\footnotesize Table \thetable}
         \def\@listI{\leftmargin\leftmargini\parsep=0pt\itemsep=0pt}
         \def\thebibliography#1{\section{References}\vspace*{-10pt}\list
          {[\arabic{enumi}]}{\settowidth\labelwidth{[#1]}\leftmargin\labelwidth
          \advance\leftmargin\labelsep
          \usecounter{enumi}}
          \def\newblock{\hskip .11em plus .33em minus .07em}
          \sloppy\clubpenalty4000\widowpenalty4000
          \sfcode`\.=1000\relax}
         
         \def\@nomath#1{\ifmmode \fi}
         \def\mmycite{\@ifnextchar [{\@tempswatrue\@mmycitex}
             {\@tempswafalse\@mmycitex[]}}
         \def\@mmycitex[#1]#2{\if@filesw\immediate%
         \write\@auxout{\string\citation{#2}}\fi
           \def\@citea{}\@mmycite{\@for\@citeb:=#2\do
             {\@citea\def\@citea{,}\@ifundefined
                {b@\@citeb}{{\bf ?}\@warning
                {Citation `\@citeb' on page \thepage \space undefined}}%
         \hbox{\csname b@\@citeb\endcsname}}}{#1}}
         \def\@mmycite#1#2{{{\scriptsize#1}\if@tempswa , #2\fi}}
         \def\mycite#1{$^{\protect\mmycite{#1}}$}
         \def\@cite#1#2{{#1\if@tempswa , #2\fi}}
         \def\thesection {\arabic{section}}
         \def\section#1{\addtocounter{section}{1}\setcounter{subsection}{0}
              \bigskip\medskip{\noindent\bf\thesection.\ #1}
              \medskip}
         \def\thesubsection {\arabic{section}.\arabic{subsection}}
         \def\subsection#1{\addtocounter{subsection}{1}
              \medskip{\noindent\thesubsection.\ #1}
              \medskip}
         \makeatother

\def\au#1 {\begin{center} #1 \end{center}}
\def\case#1/#2{{\textstyle\frac{#1}{#2}}}
\def\cntr#1 {\begin{center} #1 \end{center}}
\def\eq#1 #2 {\begin{equation} \label{#1} #2 \end{equation}}
\def\eqa#1 #2 #3 {\begin{eqnarray} \label{#1} #2 \label{#3} \end{eqnarray}}

\def\fig#1 #2 #3 #4 {\begin{figure} \vspace{#3pt} \caption[#1]{#4} \label{#1}
\end{figure}}

\def\tbl#1 #2 #3 #4 {\begin{table} \caption[#1]{#3} \label{#1} \vspace{-6pt}
\begin{center} {\begin{tabular}{#2}  \hline\hline #4 \vspace{1pt} \\
\hline\hline \end{tabular}} \end{center} \end{table}}

\def\tblb#1 #2 #3 #4 {\begin{table}[b] \caption[#1]{#3} \label{#1}
\vspace{-6pt} \begin{center} {\begin{tabular}{#2}  \hline\hline #4 \vspace{1pt}
\\ \hline\hline \end{tabular}} \end{center} \end{table}}

\def\tblh#1 #2 #3 #4 {\begin{table}[h] \caption[#1]{#3} \label{#1}
\vspace{-6pt} \begin{center} {\begin{tabular}{#2}  \hline\hline #4 \vspace{1pt}
\\ \hline\hline \end{tabular}} \end{center} \end{table}}

\def\ti#1 {\begin{center} \baselineskip=17pt {\large #1} \end{center}}

\def\tibf#1 {\begin{center} \baselineskip=17.5pt {\large \bf #1} \end{center}}

\def\aleq{\hspace{-6.712pt}&=&\hspace{-6.712pt}}
\def\alsp{\hspace{-6.712pt}& &\hspace{-6.712pt}}

\def\ess{\hskip.444444em plus .499997em minus .037036em}
\def\mss{\hskip.333333em plus .208331em minus .088889em}

\def\sen{\hbox{\scriptsize--}}
\def\eV{e\kern-.10emV }
\def\eVcm{e\kern-.10emV\kern-.15em,\mss}
\def\eVcl{e\kern-.10emV\kern-.10em:\ess}
\def\eVsc{e\kern-.10emV\kern-.10em;\mss}
\def\eVp{e\kern-.10emV\kern-.15em.\ess}
\def\eVpr{e\kern-.10emV) }
\def\eVc{e\kern-.10emV\kern-.10em/\kern-.10em$c$ }
\def\eVccm{e\kern-.10emV\kern-.10em/\kern-.10em$c$, }
\def\eVcp{e\kern-.10emV\kern-.10em/\kern-.10em$c$. }
\def\eVf{e\kern-.10emV\kern-.10em/fm }
\def\eVfcm{e\kern-.10emV\kern-.10em/fm, }
\def\eVfp{e\kern-.10emV\kern-.10em/fm. }
\def\bit{\begin{itemize}}
\def\eit{\end{itemize}}
\def\qids{\dot{q}_i^{\;2}}

\def\calP{{\cal P}}

\def\gs{{g_{\rm s}}}
\def\gv{{g_{\rm v}}}
\def\ms{{m_{\rm s}}}
\def\mv{{m_{\rm v}}}
\def\msv{{m_{\rm s,v}}}

\def\part{{\partial_t}}

\def\Ms{M_{\rm s}}
\def\Mv{M_{\rm v}}
\def\Mts{\widetilde M^*}

\begin{document}

\setcounter{totalnumber}{1}
\setcounter{topnumber}{1}
\setcounter{bottomnumber}{1}
\renewcommand{\topfraction}{1.0}
\renewcommand{\bottomfraction}{1.0}
\renewcommand{\textfraction}{0.0}

\vspace*{16pt}

\tibf{Particle-Production Mechanism\\
in Relativistic Heavy-Ion Collisions}
\au{Brian W. Bush and \underline{J. Rayford Nix}}

\cntr{Los Alamos National Laboratory\\
Los Alamos, New Mexico 87545, USA}
\vspace{4.9pt}

\begin{abstract}
We discuss the production of particles in relativistic heavy-ion collisions
through the mechanism of massive bremsstrahlung, in which massive mesons are
emitted during rapid nucleon acceleration.  This mechanism is described within
the framework of classical hadrodynamics for extended nucleons, corresponding
to nucleons of finite size interacting with massive meson fields.  This new
theory provides a natural covariant microscopic approach to relativistic
heavy-ion collisions that includes automatically spacetime nonlocality and
retardation, nonequilibrium phenomena, interactions among all nucleons, and
particle production.  Inclusion of the finite nucleon size cures the
difficulties with preacceleration and runaway solutions that have plagued the
classical theory of self-interacting point particles.  For the soft reactions
that dominate nucleon-nucleon collisions, a significant fraction of the
incident center-of-mass energy is radiated through massive bremsstrahlung.  In
the present version of the theory, this radiated energy is in the form of
neutral scalar ($\sigma$) and neutral vector ($\omega$) mesons, which
subsequently decay primarily into pions with some photons also.  Additional
meson fields that are known to be important from nucleon-nucleon scattering
experiments should be incorporated in the future, in which case the radiated
energy would also contain isovector pseudoscalar ($\pi^+$, $\pi^-$, $\pi^0$),
isovector scalar ($\delta^+$, $\delta^-$, $\delta^0$), isovector vector
($\rho^+$, $\rho^-$, $\rho^0$), and neutral pseudoscalar ($\eta$) mesons.
\end{abstract}

\section{Introduction}

Many particles are produced in a typical relativistic heavy-ion
collision.\mycite{Ta89,qm91}\ess  For the production of these particles, we
would like to discuss the mechanism of massive bremsstrahlung, in which massive
mesons are emitted during rapid nucleon acceleration.  This mechanism is
described within the framework of classical hadrodynamics for extended
nucleons, corresponding to nucleons of finite size interacting with massive
meson fields.  This approach, which satisfies {\it a priori\/} the physical
conditions that exist at relativistic energies, is manifestly Lorentz-covariant
and allows for nonequilibrium phenomena, interactions among all nucleons, and
particle production.

Although the nucleon is a composite particle made up of three valence quarks
plus additional sea quarks and gluons, when nucleons collide at very high
energies, only a few rare events correspond to the head-on or hard collisions
between the individual quarks and/or gluons.  Whereas the underlying
quark-gluon structure of the nucleon is of crucial importance for describing
particle production in hard collisions, such collisions are nevertheless
extremely rare, typically one in a billion.\mycite{Pe87}\ess  The vast majority
of events correspond to soft collisions not involving individual quarks or
gluons.  For describing particle production in such events, an appealing idea
is to regard the nucleon as a single extended object interacting with other
nucleons through the conventional exchange of mesons (whose underlying
quark-antiquark composition is ignored).

\fig nucleonf ldrd/uhit/nucleon 253 {Slice through the center of a nucleon.
The circle indicates the location of the root-mean-square radius, where the
exponentially decreasing mass density is only 3\% of its central value.}

Experiments involving elastic electron scattering off protons have determined
that the proton charge density is approximately exponential in
shape,\mycite{SBSW80}\mss with a root-mean-square radius of 0.862 $\pm$ 0.012
fm.  Although many questions remain concerning the relationship between the
proton charge density and the nucleon mass density,\mycite{Bh88}\mss it should
be a fairly accurate approximation to regard them as equal.  We therefore take
the nucleon mass density to be

\eq  rhonuc {\rho(r) = \frac{\mu^3}{8\pi} \exp(-\mu r) ~,} with $\mu =
\sqrt{12}/R_{\rm rms}$ and $R_{\rm rms}$ = 0.862 fm.  We show in
fig.~\ref{nucleonf} a gray-scale plot of the mass density through the center of
a nucleon calculated according to this exponential, with the root-mean-square
radius indicated by a circle.

The physical input underlying our new approach consists of Lorentz invariance
(which includes energy and momentum conservation), nucleons of finite size
interacting with massive meson fields, and the classical approximation applied
in domains where it should be reasonably valid.  At bombarding energies of many
G\eV per nucleon, the de~Broglie wavelength of projectile nucleons is extremely
small compared to all other length scales in the problem.  In addition, the
Compton wavelength of the nucleon is small compared to its radius, so that
effects due to the intrinsic size of the nucleon dominate those due to quantum
uncertainty.  Finally, the angular momentum is typically several hundred
$\hbar$, and the radiated energy corresponds to several meson masses.  The
classical approximation for nucleon trajectories should therefore be valid,
provided that the effect of the finite nucleon size on the equations of motion
is taken into account.

We describe in sect.~2 the present version of our theory, which includes the
neutral scalar ($\sigma$) and neutral vector ($\omega$) meson fields.  This
permits a qualitative discussion of not only particle production through
massive bremsstrahlung, but also such other physically relevant points as the
effect of the finite nucleon size on the equations of motion and an inherent
spacetime nonlocality that may be responsible for significant collective
effects.  The $\sigma$ and $\omega$ mesons that are produced will subsequently
decay primarily into pions with some photons also.  The resulting classical
relativistic equations of motion are solved in sect.~3 for soft nucleon-nucleon
collisions at $p_{\rm lab}$ = 14.6, 30, 60, 100, and 200 G\eVcp  Section~4
discusses the future incorporation of additional meson fields that are known to
be important from nucleon-nucleon scattering experiments, including the
isovector pseudoscalar ($\pi^+$, $\pi^-$, $\pi^0$), isovector scalar
($\delta^+$, $\delta^-$, $\delta^0$), isovector vector ($\rho^+$, $\rho^-$,
$\rho^0$), and neutral pseudoscalar ($\eta$).  Further details are given in a
series of papers,\mycite{SHN90}$^{\sen}$\mycite{BN94}\mss although not all of
the equations appearing in some of the earlier publications are in their final
form.
\bigskip

\section{Equations of motion}

Our action for $N$ extended, unexcited nucleons interacting with massive scalar
and vector meson fields is
\begin{eqnarray}\label{action}
I \aleq \overbrace{- M_0\sum_{i=1}^N \int d \tau_i \, \sqrt{\qids}}^{\sf
Nucleons} + \overbrace{\frac{1}{8\pi} \int d^4 \! x \left [(\partial \phi)^2 -
\ms^{\!2} \, \phi^2 \right ]}^{\sf Scalar~field} \nonumber \\ \alsp \mbox{} -
\underbrace{\frac{1}{8\pi} \int d^4 \! x \left ( \frac{1}{2} G^2 - \mv^{\!2} \,
V^2 \right )}_{\sf Vector~field} - \underbrace{\int d^4 \! x \left ( j \phi + K
\cdot V \right )}_{\sf Interaction} ~,
\end{eqnarray}
where $M_0$ is the bare nucleon mass and $q_i$ is the four-position of the
$i$th nucleon, whose trajectory is given by $q_i = q_i(\tau_i)$.  A dot
represents the derivative with respect to $\tau_i$.  In the action the
four-velocities are not constrained so that $\qids = 1$ and $\tau_i$ is not yet
identified as the proper time; it is only in the equations of motion, which are
derived as a result of the variation of $I$, that this is true.  We use the
metric $g^{\mu\nu} = {\rm diag}(1,-1,-1,-1)$, write four-vectors as $q^\mu =
(q^0, {\bf q}) = (q^t, q^x, q^y, q^z)$, and use units in which $\hbar = c =
1$.  The scalar potential is denoted by $\phi$, the four-vector potential by
$V$, and the meson masses by $\msv$.  The vector field strength tensor is
\begin{equation}\label{vfst}
G^{\mu\nu}  = \partial^{[\mu} V^{\nu]} \equiv \partial^\mu V^\nu - \partial^\nu
V^\mu ~,
\end{equation}
the scalar source density is
\begin{equation}\label{ss}
j(x)  = \gs \sum_{i=1}^N \int d \tau_i \, \rho(x - q_i, \dot{q}_i) \sqrt{\qids}
{}~,
\end{equation}
and the vector source density is
\begin{equation}\label{vs}
K^\mu(x) = \gv  \sum_{i=1}^N \int d \tau_i \, \rho(x - q_i, \dot{q}_i) \,
\dot{q}_i^\mu ~,
\end{equation}
where $\rho$ is the four-dimensional mass density of the nucleon, the spatial
part of which we assume to be exponential in the nucleon's rest frame.  The
values\mycite{SW86}$^{\sen}$\mycite{So79} that we have used here for the six
physical constants appearing in our theory are nucleon mass \linebreak $M$ $=$
938.91897 M\eVcm scalar ($\sigma$) meson mass $\ms$ $=$ 550 M\eVcm vector
($\omega$) meson mass $\mv$ $=$ 781.95 M\eVcm scalar interaction strength
$\gs^2$ $=$ 7.29, vector interaction strength \linebreak $\gv^2$ $=$ 10.81, and
nucleon r.m.s.\ radius $R_{\rm rms}$ $=$ 0.862 fm.

In ref.~\cite{BN93} we have derived exact equations of motion for the above
action in two limits:  (1) relativistic point nucleons and (2) nonrelativistic
extended nucleons.  We then generalize covariantly to obtain relativistic
equations of motion for extended nucleons, which can be written as
\begin{equation}
M^*_i a_i^\mu = f_{{\rm s},i}^\mu + f_{{\rm v},i}^\mu + f_{{\rm s,ext},i}^\mu +
f_{{\rm v,ext},i}^\mu ~.
\end{equation}
The effective mass is given by
\begin{eqnarray}
M^*_i \aleq \Mts + \Delta M_{{\rm self,}i} + \gs \bar \phi_{{\rm ext,}i} ~,
\\ \Mts \aleq M + \case 2/3 \Ms + \case 1/3 \Ms' - \case 4/3 \Mv + \case 1/3
\Mv' ~, \\
\Delta M_{{\rm self,}i} \aleq - \gs^2 \! \int_0^\infty \! d \sigma \! \left [
\bar h \! \left ( \frac{\sigma}{2} \right ) - \ms \int_0^{s'_i} d \zeta \, \bar
h \! \left ( \frac{\sqrt{s'^2_i - \zeta^2}}{2} \right ) J_1(\ms \zeta) \right ]
- 2 \Ms ~, \\
\gs \bar \phi_{{\rm ext,}i} \aleq - \frac{\gs^2}{2} \sum_{j \neq i}
\int_0^\infty \! d \sigma \left [ w \! \left ( \frac{k'_j}{2}, \sqrt{k'^2_j -
s'^2_j} \right ) - \ms \! \int_0^{k'_j} \! d \zeta \, w \! \left (
\frac{\sqrt{k'^2_j - \zeta^2}}{2}, \sqrt{k'^2_j - s'^2_j} \right ) \right .
\nonumber \\
& ~ & \left . \vphantom{\frac{\sqrt{k'^2_j - \zeta^2}}{2}} \mbox{} \times
J_1(\ms \zeta) \right ] ~.
\end{eqnarray}
The hadrostatic self-energies of the scalar and vector fields are denoted by
$M_{\rm s,v}$ and their logarithmic derivatives with respect to meson mass by
$M'_{\rm s,v}$.  The self-forces are given by
\begin{eqnarray}
f_{{\rm s},i}^\mu \aleq \frac{\gs^2}{12} \calP^{\mu\nu}_i \int_0^\infty d
\sigma \left [ h' \! \left ( \frac{\sigma}{2} \right ) - \ms \int_0^{s'_i} d
\zeta \, h' \! \left ( \frac{\sqrt{s'^2_i - \zeta^2}}{2} \right ) J_1(\ms
\zeta) \right ] s'_{i\nu} ~, \\
f_{{\rm v},i}^\mu \aleq \frac{\gv^2}{6} \calP^{\mu\nu}_i \int_0^\infty d \sigma
\, \frac{(s'_i \cdot v'_i) (s'_i \cdot v_i)}{s'^2_i} \left \{ \tilde h \! \left
( \frac{\sigma}{2} \right ) - \mv \int_0^{s'_i} d \zeta \left [ \tilde h \!
\left ( \frac{\sqrt{s'^2_i - \zeta^2}}{2} \right ) \right . \right . \nonumber
\\
& ~ & \left . \left . \mbox{} + \left ( 1 - \frac{(v'_i \cdot v_i)
s'^2_i}{(s'_i \cdot v'_i) (s'_i \cdot v_i)} \right ) h' \! \left (
\frac{\sqrt{s'^2_i - \zeta^2}}{2} \right ) \right ] J_1(\mv \zeta) \right \}
s'_{i\nu} ~,
\end{eqnarray}
and the external forces by
\begin{eqnarray}
\label{eq:fsext}
f_{{\rm s,ext},i}^\mu \aleq \frac{\gs^2}{2} {\cal P}^{\mu\nu}_i \! \sum_{j \neq
i} \int_0^\infty \! d \sigma \left [ w' \! \left ( \frac{k'_j}{2}, \sqrt{k'^2_j
- s'^2_j} \right ) - \ms \! \int_0^{k'_j} \! d \zeta \, w' \! \left (
\frac{\sqrt{k'^2_j - \zeta^2}}{2}, \sqrt{k'^2_j - s'^2_j} \right ) \right .
\nonumber \\
& ~ & \left . \vphantom{\frac{\sqrt{k'^2_j - \zeta^2}}{2}} \mbox{} \times
J_1(\ms \zeta) \right ] s'_{j\nu} ~, \\
\label{eq:fvext}
f_{{\rm v,ext},i}^\mu \aleq - \frac{\gv^2}{2} v_{i\nu} \! \sum_{j \neq i}
\int_0^\infty \!\! d \sigma \! \left [ w' \! \left ( \frac{k'_j}{2},
\sqrt{k'^2_j - s'^2_j} \right ) - \mv \! \int_0^{k'_j} \!\! d \zeta \, w' \!
\left ( \frac{\sqrt{k'^2_j - \zeta^2}}{2}, \sqrt{k'^2_j - s'^2_j} \right )
\right . \nonumber \\
& ~ & \left . \vphantom{\frac{\sqrt{k'^2_j - \zeta^2}}{2}} \mbox{} \times
J_1(\mv \zeta) \right ] s'^{[\mu}_j v'^{\nu]}_j ~.
\end{eqnarray}
In the above $\calP_i^{\mu\nu} = g^{\mu\nu} - v_i^\mu v_i^\nu$, $s'_j =
q_i(\tau_i) - q_j(\tau_j - \sigma)$, $v_j = \dot q_j(\tau_j)$, $v'_j = \dot
q_j(\tau_j - \sigma)$, $a_j = \ddot q_j(\tau_j)$, $k'_j = s'_j \cdot \dot
q'_j$, and the retarded proper time $\tau_j$ is determined implicitly from the
condition $k'_j = 0$, with $\sigma = 0$.

These equations are written in terms of the nucleon structure functions $h$ and
$w$ and quantities derived from them.  We define the interaction energy
function to be
\begin{equation}
\label{eq:wwdefi}
W(m,r) \equiv \int d^3 \! x_1 \int d^3 \! x_2 \, \rho({\bf x}_1) \frac{e^{- m
R}}{R} \rho({\bf x}_2) ~,
\end{equation}
where $R \equiv | {\bf x}_1 - {\bf x}_2  |$ and $r$ is the distance between the
centers of the two particles.  Here $\rho(r)$ is the nucleon mass density
normalized so that $4 \pi \int_0^\infty r^2 d r \rho(r) = 1$.  A Laplace
transform relates the structure function $w$ and its derivative $w'(\sigma,r)
\equiv r^{-1} \partial w / \partial r$ to the interaction energy $W$:
\begin{equation}
w(\sigma,r) \equiv 2 {\cal L}^{(-1)} [ W(\underline{m},r); 2 \sigma ] ~.
\end{equation}
The self-interaction structure function is
\begin{equation}
\label{eq:strfun}
h(\sigma) \equiv 32 \pi^2 \int_0^\infty d \sigma' \left ( \sigma'^2 - \sigma^2
\right ) \rho(\sigma + \sigma') \rho(|\sigma - \sigma'|) ~,
\end{equation}
with $h'(\sigma) \equiv d h(\sigma) / d \sigma$, $\bar h(\sigma) \equiv
\int_0^\sigma d \sigma' h(\sigma')$, and $\tilde h(\sigma) \equiv h'(\sigma) -
6 \mv^2 \bar h(\sigma)$.  The first-order Bessel function of the first kind is
denoted by $J_1$.

These equations of motion, which are second-order, nonlinear,
integrodifferential equations with four dimensions per particle, can be solved
numerically without further approximation.  In particular, we do not need to
make either a mean-field approximation, a perturbative expansion in coupling
strength, or a superposition of two-body collisions.  To solve them we use a
fourth-order Adams-Moulton predictor-corrector algorithm with adaptive step
sizes.  The integrations over proper time are done with a special
error-minimizing application of Lagrange's four-point (cubic) interpolation
formulas.
\bigskip

\section{Radiated energy and other results for soft nucleon-nucleon collisions}

We now present some results obtained by solving our equations of motion for the
soft collision of two nucleons at laboratory momentum $p_{\rm lab}$ = 14.6, 30,
60, 100, and 200 G\eVcp   At three of these momenta substantial experimental
data exist for heavy-ion collisions,\mycite{Ta89,qm91}\mss and at the remaining
two momenta experimental data exist for proton-proton
collisions.\mycite{Yo86}\ess We will concentrate our discussion here on such
physically observable quantities as scattering angle, transverse momentum, and
radiated energy in the center-of-mass system, in which frame the computations
are performed.

\fig sa weiden/sa 253 {Calculated dependence of the center-of-mass scattering
angle upon impact parameter for soft nucleon-nucleon collisions at five
incident laboratory momenta.}

As shown in fig.~\ref{sa}, for a given incident momentum, the center-of-mass
scattering angle for the dominating soft reactions described by our theory has
a maximum value at a certain impact parameter and decreases to zero for both
head-on and distant collisions.  With increasing incident momentum in this
range both the maximum angle and the impact parameter at which this maximum
occurs decrease.  For ultrarelativistic collisions this impact parameter is
approximately the distance at which the transversely dominating static vector
force for extended nucleons\mycite{BN92,BN93} has its maximum.  At the other
extreme of low incident momentum, the opposing scalar and vector forces are of
similar magnitude and give rise for small impact parameter to the more
complicated behavior of the double-dot--dashed curve in fig.~\ref{sa}.

The transverse momentum has a related behavior, as shown in fig.~\ref{pt}.
For a given incident momentum, the transverse momentum for soft reactions also
has a maximum value at a certain impact parameter and decreases to zero for
both head-on and distant collisions.  The maximum transverse momentum increases
slowly with increasing incident momentum in this range, and the impact
parameter at which this maximum occurs decreases.

\fig pt weiden/pt 253 {Calculated dependence of the transverse momentum upon
impact parameter for soft nucleon-nucleon collisions at five incident
laboratory momenta.}

The center-of-mass radiated energy per nucleon for soft reactions shown in
fig.~\ref{rad} also has a maximum value at a certain impact parameter.
However, this quantity decreases to a finite value for head-on collisions and
to zero for distant collisions.  The maximum center-of-mass radiated energy per
nucleon increases strongly with increasing incident momentum.  In the present
version of the theory, this radiated energy will be in the form of $\sigma$ and
$\omega$ mesons, which will subsequently decay primarily into pions with some
photons also.  The classical approximation is expected to be valid only when
the amount of radiated energy in the center-of-mass system exceeds the mass of
the lightest meson, which is 550 M\eVp  As seen in fig.~\ref{rad}, this
condition is well satisfied for impact parameters of physical interest at the
three highest incident momenta, but not at the two lowest incident momenta.

The qualitative behavior of these results can be understood in terms of the
nature of the external forces.  The repulsive vector force scales as the
Lorentz factor $\gamma$ in both the longitudinal and transverse directions,
whereas the attractive scalar force scales as $\gamma^2$ in the longitudinal
direction and as unity in the transverse direction.  This implies that the
vector force will dominate the transverse acceleration and the scalar force
will dominate the longitudinal acceleration.  For a given impact parameter the
scattering angle and transverse momentum will be essentially proportional to
the vector interaction strength~$\gv^2$, and the radiated energy will be
essentially proportional to $\gamma$ times the scalar interaction
strength~$\gs^2$.

\fig rad weiden/rad 253 {Calculated dependence of the center-of-mass radiated
energy per nucleon upon impact parameter for soft nucleon-nucleon collisions at
five incident laboratory momenta.}
\bigskip

\section{Future directions}

{}From nucleon-nucleon scattering experiments we know that several additional
meson fields are important and must be included for a quantitative
description:\mycite{Ma89}
\bit
\item Isovector pseudoscalar ($\pi^+$, $\pi^-$, $\pi^0$)
\item Isovector scalar ($\delta^+$, $\delta^-$, $\delta^0$)
\item Isovector vector ($\rho^+$, $\rho^-$, $\rho^0$)
\item Neutral pseudoscalar ($\eta$)
\eit
The next step in the systematic development of the theory should be the
inclusion of these additional meson fields.  Once this is done, the massive
bremsstrahlung in our theory would include pions, deltas, rhos, and etas in
addition to the sigmas and omegas that are produced in the current version.

\fig rhoeff2 snowbird/rhoeff2 253 {Effective mass density for an extended
nucleon.  The intrinsic mass density is exponential with a root-mean-square
radius of 0.862~fm.}

The effects of quantum uncertainty on the equations of motion should also be
studied and included if they are important.  This should be possible by use of
techniques analogous to those used by Moniz and Sharp for nonrelativistic
quantum electrodynamics.\mycite{MS77}  There appear in the classical
nonrelativistic equations of motion for an extended electron terms of the form
$\int_{\infty} \int_{\infty} \rho(r) \, {\cal O} \rho(r') \, d^3 \! r \, d^3 \!
r'$, where the operator ${\cal O}$ is a function of ${\bf r}$ and ${\bf r'}$.
Moniz and Sharp\mycite{MS77} have shown in nonrelativistic quantum
electrodynamics that the effect of quantum mechanics on the equations of motion
is to replace such terms by terms of the form $\int_{\infty} \int_{\infty}
\rho(r) \, {\cal O_{\rm \! eff}} \, \rho_{\rm eff}(r') \, d^3 \! r \, d^3 \!
r'$ and derivatives with respect to $\lambda$ of these terms, where $\lambda =
1/(M_0)$ is the Compton wavelength associated with the particle's bare mass and
the effective operator ${\cal O_{\rm \! eff}}$ is a function of ${\bf r}$,
${\bf r'}$, and $\lambda^2 \, {\bf {\nabla\kern-.15em_{\bf r'}}}^2$.  As
illustrated in fig.~\ref{rhoeff2}, the effective mass density for an extended
nucleon oscillates around the intrinsic mass density as a function of radial
distance from the origin.\mycite{N94}\ess

In conclusion, we have shown that classical hadrodynamics requires minimal
physical input, leads to equations of motion that can be solved numerically
without further approximation, and provides a suitable framework for describing
particle production in relativistic heavy-ion collisions through the mechanism
of massive bremsstrahlung.

This work was supported by the U.~S. Department of Energy.
\pagebreak

\end{document}